\documentclass[11pt]{article}

\makeatletter\renewcommand{\section}{\@startsection
{section}{1}{\z@}{-3.5ex plus -1ex minus
    -.2ex}{2.3ex plus .2ex}{\bf }}

\makeatletter\renewcommand{\subsection}{\@startsection{subsection}{2}{\z@}{-3.25ex
plus -1ex minus
   -.2ex}{1.5ex plus .2ex}{\it }}
\makeatletter\renewcommand{\subsubsection}{\@startsection{subsubsection}{3}{-2.45ex}{-3.25ex
plus -1ex minus -.2ex}{1.5ex plus .2ex}{\it }}

\makeatletter \@addtoreset{equation}{section}

\usepackage{graphicx}
\usepackage{epsfig}

\newcommand{\be}{\begin{equation}}
\newcommand{\ee}{\end{equation}}
\newcommand{\bea}{\begin{array}}
\newcommand{\ea}{\end{array}}
\newcommand{\beqa}{\begin{eqnarray}}
\newcommand{\eeqa}{\end{eqnarray}}
\newcommand{\nn}{\nonumber}

\renewenvironment{thebibliography}[1]
     {\baselineskip=16pt plus 2pt minus 1pt
      \section*{\large\refname
        \@mkboth{\MakeUppercase\refname}{\MakeUppercase\refname}}%
     \list{\@biblabel{\@arabic\c@enumiv}}%
           {\settowidth\labelwidth{\@biblabel{#1}}%
            \leftmargin\labelwidth
            \advance\leftmargin\labelsep
            \@openbib@code
            \usecounter{enumiv}%
            \let\p@enumiv\@empty
            \renewcommand\theenumiv{\@arabic\c@enumiv}}%
      \sloppy
      \clubpenalty4000
      \@clubpenalty \clubpenalty
      \widowpenalty4000%
      \sfcode`\.\@m}

\let\fn\footnote
\renewcommand{\footnote}[1]{\linespread{1.1}\fn{#1}\linespread{1.29}}

\usepackage{amsfonts}
\usepackage{amssymb}
\usepackage{mathrsfs}
\usepackage{amsmath,amssymb}
\usepackage{bbm}
\usepackage{bm}

\def\tyng(#1){\hbox{\tiny$\yng(#1)$}}

\begin{document}

\begin{titlepage}
\begin{flushright}
ITP-UH-10/08\\
\end{flushright}
\vskip 2.0cm

\begin{center}

\centerline{{\Large \bf Noncommutative Nonlinear Sigma Models and Integrability}}

\vskip 2em

\centerline{\large \bf Se\c{c}kin~K\"{u}rk\c{c}\"{u}o\v{g}lu}

\vskip 2em

{\small \centerline{\sl Institut f\"ur Theoretische Physik,
Leibniz Universit\"at Hannover} \centerline{\sl Appelstra\ss{}e 2,
D-30167 Hannover, Germany}

\vskip 1em

{\sl  e-mail:}  \hskip 2mm {\sl
seckin@itp.uni-hannover.de} }
\end{center}

\vskip 2cm

\begin{quote}
\begin{center}
{\bf Abstract}
\end{center}

We first review the result that the noncommutative principal
chiral model has an infinite tower of conserved currents, and
discuss the special case of the noncommutative ${\mathbb C}P^1$ model in
some detail. Next, we focus our attention to a submodel of
the ${\mathbb C}P^1$ model in the noncommutative spacetime 
${\mathcal A}_\theta({\mathbb R}^{2+1})$. By extending a
generalized zero-curvature representation to ${\mathcal A}_\theta({\mathbb
R}^{2+1})$ we discuss its integrability and construct its
infinitely many conserved currents. Supersymmetric principal
chiral model with and without the WZW term and a SUSY extension of
the ${\mathbb C}P^1$ submodel in noncommutative
spacetime [i.e in superspaces ${\mathcal A}_\theta({\mathbb
R}^{1+1 \, | 2})$, ${\mathcal A}_\theta({\mathbb R}^{2+1 \, | 2})$] 
are also examined in detail and their infinitely many conserved
currents are given in a systematic manner. Finally, we discuss the
solutions of the aforementioned submodels with or without SUSY.

\vskip 5pt
\end{quote}

\vskip 1cm
\begin{quote}
Pacs: 02.30.Ik, 02.40.Gh, 11.10.Lm, 11.10.Nx, 11.30.-j, 11.30.Pb 
\end{quote}

\end{titlepage}

\setcounter{footnote}{0}

\newpage

\section{Introduction}

Principal chiral models and several of its subfamilies such as the $O(N)$ and
the ${\mathbb C}P^N$ models, are important examples of classically integrable field theories
\cite{Pohlmeyer, Mikhailov, Shabat, Eichenherr-Forger}. These nonlinear
systems possess many interesting features due to their
integrability \cite{Faddeev, Chau0}. Among these, the existence
of a linear system of equations and of an infinite number of
conservation laws associated with nonlocal charges are two central
properties from which others (such as the B\"{a}cklund
transformations) can be obtained. Making use of the conserved,
curvature free connections present in these models, an infinite
number of conserved currents can be explicitly constructed by an
inductive procedure due to Br\'{e}zin et. al. \cite{Brezin}, and a
linear system of equations can thereby be easily obtained via
introducing a spectral parameter. It can be verified that the latter
imply the field equations as well as the zero-curvature condition
on the appropriate connection. Nonlocal charges, if conserved at the quantum level,
play a crucial role in finding the $S$-matrix and proving its factorizability, 
and hence the quantum integrability of a given model. It is known that $O(N)$ and ${\mathbb C}P^1$ models \cite{Luscher1, Zamolodchikov} and principal chiral models based on certain classical groups \cite{Ogievetsky, Balog} are quantum 
integrable, while ${\mathbb C}P^N$ ($N \geq 2$) is not \cite{Abdalla1}. More generally, sigma models
on compact symmetric spaces $G/H$ with $H$ simple are known to be quantum integrable \cite{Abdalla2}. 

Supersymmetric(SUSY) extensions of these nonlinear systems both at the classical and at the quantum level have also
been extensively studied in the past few decades \cite{Mikhailov:1978ji, Witten-Shankar, Abdalla3, Koberle, Evans-Hollowood, Zachos, Chau}. At the classical level, conserved currents of the supersymmetric $O(N)$ and ${\mathbb C}P^N$ models were derived
in component formalism in \cite{Zachos}. Later on, a much simpler superfield formulation with or without the SUSY WZW term was
given in \cite{Chau}. In \cite{Abdalla3}, it was shown that supersymmetry renders the ${\mathbb C}P^N$ model quantum integrable. 

Noncommutative(NC) field theories have been under investigation for
about a decade now. (See, for instance \cite{Nekrasov, Szabo} for comprehensive reviews) 
Among them, field theories defined on the Groenewold-Moyal (GM)-type deformations
of spacetime [i.e., the noncommutative algebra ${\mathcal A}_\theta({\mathbb R}^{(d+1)})$] 
hold a considerably large part of the literature. Formulation of
instantons and solitons in GM spacetime and other
noncommutative spaces, such as the noncommutative tori and fuzzy
spaces, has been extensively studied and found to present very
rich mathematical structures \cite{Nekrasov, Szabo, Harvey1, Fuzzy}. It has
been found out that such noncommutative deformations of extended
field configurations may be useful in studying the physics of
D-branes, as certain low energy limits in string theory in the
presence of background magnetic fields lead to noncommutative
Yang-Mills (YM) theories \cite{Witten}, \cite{Harvey2, Lechtenfeld-Popov4}.

Integrability properties of noncommutative nonlinear theories have
been under investigation in the past decade as well. In
\cite{Dimakis}, Dimakis and M\"{u}ller-Hoissen have studied the
existence and construction of conserved currents in nonlinear
sigma models on noncommutative spaces where an appropriate notion
of the Hodge operator can be prescribed, including the
GM plane. Formulation of nonlinear sigma models on
noncommutative $2-$torus with two-point target space and
construction of its conserved currents along the lines of
\cite{Brezin} were given in \cite{Dabrowski}.

In \cite{Lechtenfeld-Popov0}, a linear system of equations for noncommutative 
YM theory has been presented and it has been employed to discuss 
the construction of the NC 't Hooft instantons using the splitting approach. Later on, in \cite{Horvath}
the presence of this linear system was used to study the formulation of YM instantons 
via the dressing and splitting methods, and in \cite{Lechtenfeld-Popov1} that of monopoles by
solving the appropriate Riemann-Hilbert problem, after a dimensional reduction.
Another example of an integrable noncommutative theory is the $U(N)$ Ward
model studied in Ref. \cite{Lechtenfeld-Popov-Chu}. This model is
formulated in ${\mathcal A}_\theta({\mathbb R}^{2+1})$ and it
too explicitly exhibits a linear system implying the equation of
motion, and applying the dressing method gives a systematic way to
construct its solitonic solutions. It is worthwhile to note that,
particular noncommutative extensions of WZW and sine-Gordon models
are obtained from this system via dimensional reduction. The
latter possess several attractive features as discussed in
\cite{Lechtenfeld-Popov2} and \cite{Kurkcuoglu}. Supersymmetric
extensions of the noncommutative Ward model and its solitonic solutions
are recently considered in \cite{Lechtenfeld-Popov3}.

In this paper, our purpose is to discuss the integrability properties of
nonlinear sigma models defined on the GM spacetime. In particular, we will focus on the construction of
an infinite number of conserved currents of the
principal chiral model in ${\mathcal A}_\theta({\mathbb R}^{1+1})$, the ${\mathbb C}P^1$ model,
and a certain ${\mathbb C}P^1$ submodel in ${\mathcal A}_\theta({\mathbb R}^{2+1})$.
We will also treat their supersymmetric extensions.
In section 2, we start by describing the integrability
properties of the principal chiral model in ${\mathcal
A}_\theta({\mathbb R}^{(1+1)})$. Our presentation in section 2.1 has overlaps with the
previous investigations in \cite{Dimakis}. Then we specialize
to the NC ${\mathbb C}P^1$ model \cite{Lee}, discuss its relevant properties and
present its Noether currents explicitly.

In section 3, we focus our attention to a certain ${\mathbb C}P^1$
submodel in ${\mathcal A}_\theta({\mathbb R}^{2+1})$. A novel
approach to exploring integrability in $d+1$ dimensions was
introduced by Alvarez et. al. in \cite{Alvarez}, and it
essentially consists of formulating a generalized zero-curvature
condition by introducing a $d$-form connection. Quite
interestingly, this new formulation helps to reveal the existence
of an infinite number of conserved quantities in a variety of models, such as those found
for a submodel of ${\mathbb C}P^1$ model in $2+1$ dimensions. By extending
this approach and a parallel one developed by Fujii et. al.
\cite{Fujii} to noncommutative spacetime, we discuss the
integrability properties of the aforementioned ${\mathbb C}P^1$ submodel and
construct an infinite number of conserved currents for it in a
systematic manner. We also discuss the solitonic solutions of the
submodel in some detail and show that BPS solutions of the NC
${\mathbb C}P^1$ model are solutions of the submodel too.

In section 4, we examine the supersymmetric principal chiral model in 
${\mathcal A}_\theta({\mathbb R}^{1+1 \, | 2})$
with and without the WZW term in some detail. We discuss the
integrability properties of these models and derive their
conserved currents in the superfield formalism, using the methods
of \cite{Chau}. This is followed by a study of the SUSY extension of
the ${\mathbb C}P^1$ submodel in ${\mathcal A}_\theta({\mathbb
R}^{(2+1) \, | 2})$ and construction of its conserved currents.
Solitonic configurations of this model are also given.
We conclude by summarizing our results and stating some
directions we are going to be exploring in the near future.

Until section 4, we will be working on the noncommutative
spacetimes ${\mathcal A}_\theta({\mathbb R}^{1+1})$ and
${\mathcal A}_\theta({\mathbb R}^{2+1})$, which are defined by
the commutation relations
\be
\lbrack {\hat x}_\mu \,, {\hat x}_\nu \rbrack = i \theta_{\mu \nu}\,,
\ee
and the indices run over $0,1$ and $0,1,2$, respectively. We use 
the Minkowski metric with signature $(+ \,,- \,, -)$. From section 4 onward,
appropriate Grassmann variables will be introduced to obtain the 
superspaces ${\mathcal A}_\theta({\mathbb R}^{1+1 \, | 2})$ and
${\mathcal A}_\theta({\mathbb R}^{2+1 \, | 2})$, where only the bosonic
coordinates do not commute.

\section{Nonlinear Models and Integrability}

\subsection{Principal Chiral Model in ${\mathcal A}_\theta({\mathbb R}^{1+1})$:}

Let us start our discussion by considering the principal chiral model in 
${\mathcal A}_\theta({\mathbb R}^{1+1})$. It is defined by the action
\be
S_{PC} = \frac{1}{4} \pi \theta \mbox{Tr} \partial_\mu g \partial^\mu g^{-1} \,,
\label{eq:pcm}
\ee
where $g$ is a nonsingular matrix whose entries are operators in ${\mathcal A}_\theta({\mathbb R}^{1+1})$ acting
on the standard Heisenberg-Weyl Hilbert space ${\cal H}$.\footnote{Note that ${\cal H}$ can not be taken in the Fock
basis due to the Minkowski signature.} For definiteness, we take $g \in U(N)$, thus it satisfies $gg^\dagger = g^\dagger g =1 $. We have that $\mbox{Tr} = \mbox{Tr}_{\cal H} \otimes \mbox{Tr}_N$, where $\mbox{Tr}_N$ is the trace in $Mat(N)$.

The equation of the motion following from $S_{PC}$ is
\be
\partial^\mu (g^{-1} \partial_\mu g) = 0 \,,
\label{eq:eompcm}
\ee
and readily implies
\be
A^{Noether}_\mu = g^{-1} \partial_\mu g \,,
\label{eq:Noether}
\ee
as the conserved Noether currents of the model under the global $U(N)$ symmetry.

To construct the conserved tower of currents, we closely follow the inductive procedure of \cite{Brezin}.
Let us first define the covariant derivative $D_\mu = \partial_\mu + A_\mu$. Due to
(\ref{eq:Noether}), it satisfies
\be
\lbrack D_\mu \,, D_\nu \rbrack = 0 \,,
\label{eq:prop1}
\ee
and due to (\ref{eq:eompcm}), we further have
\be
\partial_\mu D^\mu = D^\mu \partial_\mu \,.
\label{eq:prop2}
\ee

Let us now suppose that we have found the conserved current $J_\mu^{(n)}$ at level $n$.
By Hodge decomposition of differential forms, which applies in the present NC spacetime
${\mathcal A}_\theta({\mathbb R}^{(1+1)})$ as the algebra of derivatives are not deformed (i.e. derivatives commute),
this implies that we can find $\chi^{(n)} \in {\mathcal A}_\theta({\mathbb R}^{1+1}) \otimes Mat(N)$ such that
\be
J_\mu^{(n)} = - \epsilon_{\mu \nu} \partial^\nu \chi^{(n)} \,, \quad  n \geq 1 \,,
\label{eq:def1}
\ee
Then, the $(n+1)^{th}$ current is
\be
J_\mu^{(n+1)} = D_\mu \chi^{(n)} \,, \quad n \geq 0.
\label{eq:def2}
\ee
The construction starts with $\chi^{(0)} = 1$ and $J_\mu^{(1)} = A^{Noether}_\mu$.
We can see that $J_\mu^{(n+1)}$ is conserved since
\beqa
\partial^\mu J_\mu^{(n+1)} &=& D_\mu \partial^\mu \chi^{(n)} \,, \quad n \geq 1 \nn \\
&=& \epsilon_{\mu \nu} D_\mu J_\nu^{(n)} \nn \\
&=& \epsilon_{\mu \nu}  D_\mu D_\nu \chi^{(n-1)} = 0 \,.
\eeqa
where we have used (\ref{eq:prop1}), (\ref{eq:prop2}) and (\ref{eq:def1}). Thus, the construction of \cite{Brezin} works 
for the noncommutative principal chiral model too. As we have already stated in the introduction, this result overlaps with
that of \cite{Dimakis}.

The form of the conserved currents allows us to define the linear system of equations for this model.
Introducing a spectral parameter $\lambda$ via $\chi = \sum_0^\infty \lambda^{-n} \chi^n$, we can write using
(\ref{eq:def1}) and (\ref{eq:def2}) that
\be
- \epsilon_{\mu \nu} \partial^\nu \chi = \lambda^{-1} D_\mu \chi \,.
\ee
The last equation can be brought into the form
\beqa
-\partial_1 \chi \, &=& \frac{\lambda A_0 + A_1}{1 - \lambda^2} \chi \,, \nn \\
-\partial_0 \chi \, &=& \frac{\lambda A_1 + A_0 }{1 - \lambda^2} \chi \,.
\label{eq:Laxpairs}
\eeqa
Obviously, the system of equations in (\ref{eq:Laxpairs}) is of the same form as that of the commutative model.
However, we note that $A_\mu({\hat x}_\mu)$, $\chi({\hat x}_\mu \,, \lambda)$ are operators in 
${\mathcal A}_\theta({\mathbb R}^{1+1}) \otimes Mat(N)$ acting on the Hilbert space
${\cal H} \otimes {\mathbb C}^N$. Solvability of the system implies the equation of motion (\ref{eq:eompcm}) and the zero-curvature condition (\ref{eq:prop1}).

The explicit form of the currents $J_\mu^{(n)}$ do indeed differ from those of the commutative model. In the following 
subsection, we present an example, namely the Noether currents of the ${\mathbb C}P^1$ model to emphasize this point.

\subsection{NC ${\mathbb C}P^1$ Model:}

We can now focus on the NC ${\mathbb C}P^1$ model \cite{Lee}. Restricting to the subset of operators of the form
\be
g = g^{-1} = e^{i \pi P} = 1 - 2P \,,
\ee
where $P$ is a projector in $ {\mathcal A}_\theta({\mathbb R}^{1+1}) \otimes Mat(2)$:
\be
P^2 = P \,, \quad P^\dagger = P \,, \quad P \in {\mathcal A}_\theta({\mathbb R}^{(1+1)}) \otimes Mat(2) \,,
\label{eq:projector1}
\ee
leads to the ${\mathbb C}P^1$ model action
\be
S = \pi \theta \mbox{Tr} \partial_\mu P \partial^\mu P \,, \quad \mu = 0,1 \,.
\label{eq:CPN}
\ee
The Noether currents take the form
\be
J^{Noether}_\mu = \lbrack P \,, \partial_\mu P \rbrack \,.
\label{eq:NoetherCP}
\ee
Let us parametrize the projector as
\be
P =
\left (
\begin{array}{cc}
\frac{1}{u^\dagger u + 1} &  \frac{1}{u^\dagger u + 1} u^\dagger \\
u \frac{1}{u^\dagger u + 1} & u \frac{1}{u^\dagger u + 1} u^\dagger
\end{array}
\right ) \,,
\label{eq:projectorlocal}
\ee
then the conservation of $J^{Noether}_\mu$ implies the field equation for $u$  
\be
\partial_\mu \partial^\mu u - 2 \partial_\mu u \frac{1}{u^\dagger u + 1} u^\dagger \partial_\mu u = 0 \,.
\ee

Using (\ref{eq:projectorlocal}), the Noether currents associated with the global $SU(2)$ symmetry take the form
\begin{multline}
J_{\mu, 3} = \frac{1}{2} \mbox{Tr}_{2} \lambda_3 \lbrack P \,, \partial_\mu P \rbrack
= \frac{1}{2} \Big ( \frac{1}{u^\dagger u + 1} (u^\dagger \partial_\mu u - \partial_\mu u^\dagger u) \frac{1}{u^\dagger u + 1}
- u \frac{1}{(u^\dagger u + 1)^2} \partial_\mu u^\dagger \\
+ \partial_\mu u \frac{1}{(u^\dagger u + 1)^2} u^\dagger - u \left \lbrack \frac{1}{u^\dagger u + 1} \,, 
\partial_\mu \frac{1}{u^\dagger u + 1} \right \rbrack u^\dagger
- \left \lbrack \frac{u^\dagger u}{u^\dagger u + 1} \,, \partial_\mu \left ( u \frac{1}{u^\dagger u + 1} u^\dagger \right )
\right \rbrack \Big ) \,,
\label{eq:Noether1}
\end{multline}
\be
J_{\mu, +} =  \frac{1}{2} \mbox{Tr}_{2} \lambda_+ \lbrack P \,, \partial_\mu P \rbrack =
- \frac{1}{2} \Big ( \partial_\mu u \frac{1}{(u^\dagger u + 1)} + u \frac{1}{u^\dagger u + 1} 
(\partial_\mu u^\dagger  u  - u^\dagger \partial_\mu u ) \frac{1}{u^\dagger u + 1} \Big ) \,,
\label{eq:Noether2}
\ee
\be
J_{\mu, -} =  \frac{1}{2} \mbox{Tr}_{2} \lambda_- \lbrack P \,, \partial_\mu P \rbrack = - J_{\mu, +}^\dagger \,.
\label{eq:Noether3}
\ee
where $\lambda_{i} \,, (i= 1,2,3)$ are the Pauli matrices and $\lambda_\pm = \lambda_1 \pm i \lambda_2$. Unlike the commutative ${\mathbb C}P^1$ model, the Noether current associated with the global $U(1)$ symmetry of the action is not zero, but it is given by
\begin{multline}
J_{\mu, 0} =  \frac{1}{2} \mbox{Tr}_{2} \lbrack P \,, \partial_\mu P \rbrack
= \frac{1}{2} \Big( \frac{1}{u^\dagger u + 1} (u^\dagger \partial_\mu u - \partial_\mu u^\dagger u) \frac{1}{u^\dagger u + 1}
+ u \frac{1}{(u^\dagger u + 1)^2} \partial_\mu u^\dagger \\
- \partial_\mu u \frac{1}{(u^\dagger u + 1)^2} u^\dagger + u \left \lbrack \frac{1}{u^\dagger u + 1} \,, 
\partial_\mu \frac{1}{u^\dagger u + 1} \right \rbrack u^\dagger
+ \left \lbrack \frac{u^\dagger u}{u^\dagger u + 1} \,, \partial_\mu \left ( u \frac{1}{u^\dagger u + 1} u^\dagger \right )
\right \rbrack \Big ) \,,
\label{eq:Noether4}
\end{multline}
In the commutative limit the standard expressions for the Noether currents are recovered. In particular,
$J_{\mu, 0}$ becomes zero in this limit.

\section{A ${\mathbb C}P^1$ submodel in ${\mathcal A}_\theta({\mathbb R}^{2+1})$}

A valuable approach to exploring integrability in $2+1$ and higher dimensional theories is due to Alvarez et. al.
\cite{Alvarez}. In this article, a generalized zero-curvature representation consisting of an appropriate curvature
free connection together with a covariantly conserved vector field has been formulated. The generalized zero-curvature
representation implies the presence of conserved currents which may be obtained in a systematic manner. In several diverse models of interest admitting this representation, it has been found that the conserved currents are infinite in number leading to their integrability. 
For instance, in certain submodels of the principal chiral models and ${\mathbb C}P^N$ models in $2+1$ dimensions,
which are determined by the requirement of additional equations to be satisfied by the fields over and above the equations of motions 
of their respective parent models, an infinite tower of conserved currents has been obtained explicitly using the generalized zero 
curvature representation \cite{Alvarez, Fujii0, Sanchez, Ferreira}. In another example in $3+1$ dimensions considered by Aratyn et. al. \cite{Aratyn1}, a full field theory possesing toroidal solitonic solutions has been shown to be integrable using the generalized zero-curvature representation and its conserved currents have been constructed.

A parallel approach to that of \cite{Alvarez} has been developed by Fujii et. al. \cite{Fujii}. In this formulation, for instance the ${\mathbb C}P^N$ submodels are studied by implementing their defining conditions as additional equations to be satisfied
by the projectors of the ${\mathbb C}P^N$ models, rather than on their particular parametrizations. This approach appears to be better suited for adapting to the present setting of noncommutative theories and will be followed in this section. However, before doing so, it seems instructive to briefly sketch how the ideas of \cite{Alvarez} fit into the current framework, and state the type of limitation it faces, in providing explicit expressions for the conserved quantities.

Suppose that we have a finite-dimensional non-semi-simple Lie algebra ${\hat {\underbar G}}$. Then we can write 
${\hat {\underbar G}} = {\underbar G} + {\cal I}$ where ${\underbar G}$ is a semisimple Lie subalgebra of ${\hat {\underbar G}}$ 
and ${\cal I}$ is its maximal solvable ideal (i.e. radical). We can consider now a connection one-form $A_\mu$
on ${\mathcal A}_\theta({\mathbb R}^{2+1})$ valued in ${\underbar G}$, and an antisymmetric tensor $B_{\mu \nu}$ 
valued in ${\cal I}$. In $2+1$ dimensions we can write the dual of $B_{\mu \nu}$ as
\be
{\tilde B}^\mu = \frac{1}{2} \varepsilon^{\mu \nu \rho} B_{\nu \rho} \,.
\ee
A generalized set of integrability conditions can then be given as \cite{Alvarez}
\be
F_{\mu \nu} = \lbrack D_\mu \,, D_\nu \rbrack = 0 \,, \quad D_\mu {\tilde B}^\mu = 0 \,, \quad D_\mu = \partial_\mu + A_\mu \,.
\ee
Since $A_\mu$ is a flat connection we can write
\be
A = g^{-1} \partial_\mu g \,, \quad g \in G
\ee
where $G$ is the Lie group whose Lie algebra is {\underbar G}. From these considerations, 
it is easy to verify that the currents
\be
J_\mu = g^{-1} {\tilde B}_\mu g \,,
\ee
are conserved. To construct these currents explicitly in a model with say $G \equiv SU(2)$, one essentially
needs a suitable local parametrization of $SU(2)$. (See. for instance. the construction of the ${\mathbb C}P^1$ submodel currents
in commutative space given in \cite{Alvarez}.) However, such a parametrization of $SU(2)$ does not exist in the noncommutative
setting, and thus the above construction remains implicit for the currents.

Let us now turn to applying the methods of \cite{Fujii}, and to be more concrete 
consider a ${\mathbb C}P^1$ submodel in ${\mathcal A}_\theta({\mathbb R}^{2+1})$.
With $P \in {\mathcal A}_\theta({\mathbb R}^{2+1}) \otimes Mat(2)$ we observe that 
the tensor product [over ${\mathcal A}_\theta({\mathbb R}^{2+1})$] $P \otimes P$ is a projector in
${\mathcal A}_\theta({\mathbb R}^{2+1}) \otimes Mat(2^2)$. Then the submodel we
are interested in may be specified by the equation \cite{Fujii}
\be
\lbrack P \otimes P \,, \partial_\mu \partial^\mu P \otimes P \rbrack = 0 \,, \quad \mu = 0,1,2 \,.
\label{eq:submodel2}
\ee
In (\ref{eq:submodel2}) and what follows the derivatives on $k$-fold tensor products are given via
\be
\partial_\mu \equiv \sum_i^{k-1} \underbrace{ 1 \otimes 1 \otimes \cdots \otimes }_i
\partial_\mu \otimes \underbrace{ 1 \otimes 1 \cdots \otimes 1}_{k-1-j} \,,
\ee
and the same symbol is used in the tensor product space, as there is no risk of confusion.

It is easy to find that (\ref{eq:submodel2}) can be expressed as the two equations
\be
\lbrack P \,, \partial_\mu \partial^\mu P \rbrack = 0  \,, 
\label{eq:eom2+1}
\ee
\be
\partial^\mu P \otimes \lbrack P \,, \partial_\mu P
\rbrack + \lbrack P \,, \partial_\mu P \rbrack \otimes \partial^\mu P = 0 \,.
\label{eq:submodel1}
\ee
Clearly, the first of these is the equation of motion for the ${\mathbb C}P^1$ model, while (\ref{eq:submodel1})
puts further restrictions on the projector $P$ and thereby specifies a submodel. Using (\ref{eq:projectorlocal}), we may also
express these conditions as
\be
\partial_\mu \partial^\mu u - 2 \partial_\mu u \frac{1}{u^\dagger u + 1} u^\dagger \partial_\mu u = 0 \,, \quad
\partial_\mu u \frac{1}{(u^\dagger u + 1)^2} u^\dagger \partial_\mu u = 0 \,.
\ee
In the commutative limit these equations collapse to $\partial^\mu \partial_\mu u = 0$ and $\partial_\mu u \partial^\mu u = 0$,
which define the submodel in the commutative space \cite{Alvarez}.

\subsection{Conserved Currents:}

In close analogy to the commutative model \cite{Fujii}, the conserved matrix currents in this model can now be constructed.
They are given by
\be
J_\mu^k = \sum_{i=0}^{k-1} \underbrace{P \otimes P \cdots \otimes P}_i \otimes \lbrack P \,, \partial_\mu P
\rbrack \otimes \underbrace{P \otimes P \cdots \otimes P}_{k-1-i} \,.
\label{eq:currents1}
\ee
It follows from (\ref{eq:eom2+1}) and (\ref{eq:submodel1}) that $J_\mu^k$ is conserved:
\be
\partial^\mu J_\mu^k = 0 \,.
\label{eq:conserve}
\ee
For instance, at level $k=3$ we have
\begin{multline}
\partial^\mu J_\mu^{k=3} = \partial^\mu \lbrack P \,, \partial_\mu P \rbrack \otimes P \otimes P + 
\partial^\mu P \otimes \lbrack P \,, \partial_\mu P \rbrack \otimes P
+ \lbrack P \,, \partial_\mu P \rbrack \otimes \partial^\mu P \otimes P \\
+ P \otimes \partial^\mu \lbrack P \,, \partial_\mu P \rbrack \otimes P
+ P \otimes \partial^\mu P \otimes \lbrack P \,, \partial_\mu P \rbrack +
P \otimes \lbrack P \,, \partial_\mu P \rbrack \otimes \partial^\mu P  \\
+  P \otimes P \otimes \partial^\mu \lbrack P \,, \partial_\mu P \rbrack +
\lbrack P \,, \partial_\mu P \rbrack \otimes P \otimes \partial_\mu P
+ \partial^\mu P \otimes P \otimes \lbrack P \,, \partial_\mu P \rbrack 
= 0 \,.
\end{multline}
upon using (\ref{eq:eom2+1}) and (\ref{eq:submodel1}).

A few simple comments are in order. Clearly, level $k=1$ in the above construction corresponds to the NC
${\mathbb C}P^1$ model and from (\ref{eq:currents1}) we recover the Noether currents of the model, as given in (\ref{eq:Noether1}, 
\ref{eq:Noether2}, \ref{eq:Noether3}, \ref{eq:Noether4}), where now the index $\mu$ in these equations run from $0$ to $2$.
Next, we observe that all the results above go through for the NC ${\mathbb C}P^N$ model, once $Mat(2^2)$ is replaced by $Mat((N+1)^2)$.
We can ask, how many conserved currents are there at a given level $k$? For the ${\mathbb C}P^1$ model, we have four conserved currents
at level $k=1$, and $2^k \times 2^k$ conserved current at level $k$, and for the ${\mathbb C}P^N$ model we have $(N+1)^k \times (N+1)^k$
conserved currents at level $k$. Clearly, the number of conserved currents tends to infinity as $k$ does so. 

A fast way to compute the component currents is to take the trace of the product of $J_\mu^k$ with elements of a
suitably chosen basis. Let us illustrate this for the simplest case $k = 2$. In this case the tensor product space is
$Mat(4)$ and it can be spanned by the basis 
\be
\Lambda_{ab} = \lambda_a \otimes \lambda_b \,, \quad \lambda_a = (1_{2}\,, \lambda_+ \,, \lambda_- \,, \lambda_3). 
\label{eq:basis}
\ee
Using the identity $\mbox{Tr} A \otimes B = \mbox{Tr} A \, \mbox{Tr} B$, we can write
\beqa
(J_\mu^{k=2})_{ab} &=& \mbox{Tr}_{4} \Lambda_{ab} J_\mu^{k=2} \nn \\
&=& \mbox{Tr}_{4} \lambda_a \otimes \lambda_b \left ( \lbrack P \,, \partial_\mu P \rbrack \otimes P +
P \otimes \lbrack P \,, \partial_\mu P \rbrack \right ) \nn \\ 
&=& \mbox{Tr}_{2} \lambda_a \lbrack P \,, \partial_\mu P \rbrack \, 
\mbox{Tr}_{2} \lambda_b P +  \mbox{Tr}_{2} \lambda_a P \,
\mbox{Tr}_{2} \lambda_b \lbrack P \,, \partial_\mu P \rbrack \,. 
\label{eq:tracetensor}
\eeqa
The $16$ conserved currents present at this level can be obtained from (\ref{eq:tracetensor}). 
We list a few examples for concreteness:
\begin{multline}
(J_\mu^{k=2})_{++} = - \Big ( \partial_\mu u \frac{1}{(u^\dagger u + 1)} + u \frac{1}{u^\dagger u + 1} 
(\partial_\mu u^\dagger  u  - u^\dagger \partial_\mu u ) \frac{1}{u^\dagger u + 1} \Big ) u \frac{1}{u^\dagger u + 1}  \\
- u \frac{1}{u^\dagger u + 1} \Big ( \partial_\mu u \frac{1}{(u^\dagger u + 1)} + u \frac{1}{u^\dagger u + 1} 
(\partial_\mu u^\dagger  u  - u^\dagger \partial_\mu u ) \frac{1}{u^\dagger u + 1} \Big ) \,,
\end{multline}
\begin{multline}
(J_\mu^{k=2})_{+-} = - \Big ( \partial_\mu u \frac{1}{(u^\dagger u + 1)} + u \frac{1}{u^\dagger u + 1} 
(\partial_\mu u^\dagger  u  - u^\dagger \partial_\mu u ) \frac{1}{u^\dagger u + 1} \Big ) \frac{1}{u^\dagger u + 1} u^\dagger \\
+ u \frac{1}{u^\dagger u + 1} \Big ( \frac{1}{u^\dagger u + 1} \partial_\mu u^\dagger - \frac{1}{u^\dagger u + 1}
(\partial_\mu u^\dagger  u  - u^\dagger \partial_\mu u ) \frac{1}{u^\dagger u + 1} u^\dagger \Big ) \,,
\end{multline}
\begin{multline}
(J_\mu^{k=2})_{+3} = - \Big ( \partial_\mu u \frac{1}{(u^\dagger u + 1)} + u \frac{1}{u^\dagger u + 1} 
(\partial_\mu u^\dagger  u  - u^\dagger \partial_\mu u ) \frac{1}{u^\dagger u + 1} \Big ) \Big(  
\frac{1}{u^\dagger u + 1} - u \frac{1}{u\dagger u + 1} u^\dagger \Big ) \\
+ u \frac{1}{u^\dagger u + 1} \Big ( \frac{1}{u^\dagger u + 1} (u^\dagger \partial_\mu u - \partial_\mu u^\dagger u) \frac{1}{u^\dagger u + 1} - u \frac{1}{(u^\dagger u + 1)^2} \partial_\mu u^\dagger \\
+ \partial_\mu u \frac{1}{(u^\dagger u + 1)^2} u^\dagger - u \left \lbrack \frac{1}{u^\dagger u + 1} \,, 
\partial_\mu \frac{1}{u^\dagger u + 1} \right \rbrack u^\dagger - 
\left \lbrack \frac{u^\dagger u}{u^\dagger u + 1} \,, \partial_\mu \left ( u \frac{1}{u^\dagger u + 1} u^\dagger \right )
\right \rbrack \Big ) \,.
\end{multline}

\subsection{Solutions of the Submodel:}

Static solitonic solutions of the noncommutative ${\mathbb C}P^1$ model are given by the BPS configurations \cite{Lee}.
In the complex coordinates $z= \frac{{\hat x}_1 + i {\hat x}_2}{\sqrt{2}}$ satisfying
$\lbrack z \,, {\bar z} \rbrack = \theta$ the BPS configurations are specified by the equations
\be
\partial_{\bar z} P P = 0 \quad \mbox{(self-dual)} \,, \quad \partial_z P P = 0 \quad \mbox{(anti-self-dual)} \,,
\label{eq:selfdual1}
\ee
where the derivatives are given by the adjoint actions
\be
\partial_z = - ad \, {\bar z} = - \lbrack {\bar z} \,, \cdot \rbrack \,,
\quad \partial_{\bar z} = ad \, z = \lbrack z \,, \cdot \rbrack \,.
\ee
In view of the fact that $\partial P = \partial P P + P \partial P$, (\ref{eq:selfdual1}) can
also be expressed in the form
\be
(1 - P) \partial_{\bar z} P = 0 \quad \mbox{(self-dual)} \,, \quad (1 - P) \partial_z P = 0 \quad \mbox{(anti-self-dual)} \,,
\label{eq:selfdual2}
\ee
Parametrizing the projector as in (\ref{eq:projectorlocal}), it can be inferred that these equations are
fulfilled by the functions $u=u(z)$ (self-dual) and $u=u({\bar z})$ (anti-self-dual) analytic in their arguments. 

Let us now show that, these configurations are also solutions of the ${\mathbb C}P^1$ submodel. Equation
(\ref{eq:eom2+1}), being the quadratic field equation for the ${\mathbb C}P^1$ model, is automatically satisfied
by $P$ fulfilling either of the two equations in (\ref{eq:selfdual1}). As for (\ref{eq:submodel1}) taking
for instance the anti-self-dual configurations we have
\be
(\ref{eq:submodel1}) = - \partial_z P \otimes \partial_{\bar z} P P + \partial_{\bar z} P
\otimes P \partial_z P + P \partial_z P \otimes \partial_{\bar z} P - \partial_{\bar z} P P \otimes \partial_z P \,,
\ee
and it vanishes identically upon using the second equation in (\ref{eq:selfdual1}) and its Hermitian conjugate.
Clearly, a similar calculation holds for the self-dual solution too.

\section{Supersymmetric Nonlinear Models}

\subsection{Noncommutative SUSY Principal Chiral Model:}

Let us now focus our attention to the ${\cal N} =1$ superspace ${\mathcal A}_\theta({\mathbb R}^{1+1 \, | 2})$ 
with Moyal-type noncommutativity, i.e.
\be
\lbrack {\hat x}_\mu \,, {\hat x}_\nu \rbrack = i \theta_{\mu \nu} \,, \quad \lbrace \theta_\alpha \,, \theta_\beta \rbrace = 0  
\,, \quad
\lbrack {\hat x}_\mu \,, \theta_\alpha \rbrack = 0 \,, \quad  \mu \,, \nu = 0,1 \,, \quad \alpha \,, \beta = 1,2 \,.
\ee
The supersymmetric principal chiral model is given by the action
\be
S = \frac{1}{4} \pi \theta  \int \, d^2 \theta \, \mbox{Tr} \, {\bar D} G^\dagger D G \,,
\ee
where the SUSY covariant derivative is
\be
D_\alpha = \frac{\partial}{\partial {\bar \theta}^\alpha } + i (\gamma^\mu \theta)_\alpha \partial_\mu
\label{eq:SUSYaction}
\ee
and $G = G(x_\mu , \theta_\alpha)$ is a matrix valued superfield in NC space with $G G^\dagger = 1 = G^\dagger G$.
For definiteness we will assume that $G \in U(N)$.

For the $\gamma$ matrices we take
\be
\gamma^0 = \left (
\begin{array}{cc}
0 & 1 \\
-1 & 0
\end{array}
\right ) \,, \quad
\gamma^1 = \left (
\begin{array}{cc}
0 & 1 \\
1 & 0
\end{array}
\right ) \,, \quad
\gamma^5 = \gamma^1 \gamma^2 = \left (
\begin{array}{cc}
1 & 0 \\
0 & -1
\end{array}
\right ) \,.
\ee

It may be noted that $\lbrace D_1 \,, D_2 \rbrace = 0$, and the commutators of $D_\alpha$ with the generators of Poincar\'{e}
algebra are the same as those without noncommutativity, therefore the full SUSY algebra is present and is undeformed.

We will now demonstrate that this model satisfies a zero-curvature condition and is therefore integrable at the classical level
and construct its conserved nonlocal currents. Our approach is the superspace generalization of that of \cite{Brezin} and was used by
Chau and Yen \cite{Chau} to construct the nonlocal charges in SUSY principal chiral models with or without the WZW term.

The equation of motion that follows from the variation of (\ref{eq:SUSYaction}) is
\be
{\bar D} (G^\dagger D G ) = 0  \,.
\label{eq:SUSYeom}
\ee
Let us define a gauge superfield as ${\cal A}_\alpha = G^\dagger D_\alpha G$. Then (\ref{eq:SUSYeom}) becomes
\be
D_1 {\cal A}_2 - D_2 {\cal A}_1 = 0 \,.
\label{eq:SUSYeom1}
\ee
Furthermore, we have the gauge covariant derivative ${\cal D}_\alpha = D_\alpha + {\cal A}_\alpha$, which immediately leads to 
zero-curvature for ${\cal A}_\alpha$:
\be
\lbrace {\cal D}_1 \,, {\cal D}_2 \rbrace = D_1 {\cal A}_2 + D_2 {\cal A}_1 + \lbrace {\cal A}_1 \,, {\cal A}_2 \rbrace = 0 \,.
\label{eq:zerocurv}
\ee
This condition together with (\ref{eq:SUSYeom}) implies that the model is integrable. As a consequence of
the equation of motion the identity
\be
\lbrace D_\alpha \,, {\bar {\cal D}}_\alpha \rbrace = 0 \,
\ee
holds. 

It is now easy to construct the nonlocal conserved currents. Suppose that we have found the
conserved current ${\cal J}_\alpha^{(n)}$ at level $n$. This implies that we can find $\xi^{(n)}
\in {\mathcal A}_\theta({\mathbb R}^{1+1 \, |2}) \otimes Mat(N)$ such that
\be
{\cal J}_1^{(n)} = - D_1 \xi^{(n)} \,, \quad {\cal J}_2^{(n)} = D_2 \xi^{(n)} \,.
\label{eq:SUSYcurrents}
\ee
Then, the $(n+1)^{th}$ current is
\be
{\cal J}_\alpha^{(n+1)} = {\cal D}_\alpha \xi^{(n)} \,, \quad n \geq 0.
\label{eq:nthcurrent}
\ee
The construction starts with $\xi^{(0)} = 1$ and ${\cal J}_\alpha^{(1)} = {\cal A}_\alpha$.
We can see that ${\cal J}_\alpha^{(n+1)}$ is conserved
\beqa
&&D_1 {\cal J}_2^{(n+1)} - D_2 {\cal J}_1^{(n+1)} \nn \\
&&= D_1 {\cal D}_2 \xi^{(n)} - D_2 {\cal D}_1 \xi^{(n)} \nn \\
&&= -{\cal D}_2 D_1 \xi^{(n)} + {\cal D}_1 D_2 \xi^{(n)} \nn \\
&&= {\cal D}_2 {\cal J}_1^{(n)} + {\cal D}_1 {\cal J}_2^{(n)} \nn \\
&&= {\cal D}_2  {\cal D}_1 \xi^{(n-1)} + {\cal D}_1 {\cal D}_2 \xi^{(n-1)} \nn \\
&&= \lbrace {\cal D}_1 \,, {\cal D}_2 \rbrace \xi^{(n-1)} = 0 \,.
\eeqa

Introducing a spectral parameter $\kappa$ and writing $\xi = \sum_n \kappa^n \xi^{(n)}$ with $\xi^{(0)} = 1$, we
find from (\ref{eq:SUSYcurrents}) and (\ref{eq:nthcurrent}) that
\beqa
D_1 \xi &=& - \frac{\kappa}{1 + \kappa} {\cal A}_1 \xi \,, \nn \\
D_2 \xi &=& \frac{\kappa}{1 - \kappa} {\cal A}_2 \xi
\eeqa 
which is precisely of the same form as in the commutative space, but now ${\cal A}_\alpha$ and $\xi$ are operators in
${\mathcal A}_\theta({\mathbb R}^{1+1 \, |2}) \otimes Mat(N)$.
 
\subsection{Addition of the WZW term:}

The supersymmetric WZW term is of the form \cite{Rossi, Abdalla} 
\be
S_{WZW} = \frac{k}{16 \pi} 2 \pi \theta \int d^2 \theta \, d t \, \mbox{Tr} \, G^\dagger 
\frac{d G}{d t } {\bar D} G^\dagger \gamma_5 D G \,,
\ee
where $k \in {\mathbb Z}$. The variation of the total action $S = S_{PC} + S_{WZW}$ yields
\be
{\bar D} \left ( (1 + \frac{k}{\pi} \gamma_5) G^\dagger D G \right ) = 0.  
\ee
We observe that all the results of the previous section hold, if we make the substitution
\be
{\cal A}_\mu \longrightarrow (1 - \frac{k}{\pi}) {\cal A}_\mu \,.
\ee
Thus, we conclude that all the classical integrability properties are possessed by the NC supersymmetric WZW model too.

\subsection{SUSY ${\mathbb C}P^1$ Model:}

The SUSY ${\mathbb C}P^1$ on ${\mathcal A}_\theta({\mathbb R}^{1+1 \, | 2})$  model is specified by
\be
G = e^{i \pi {\cal P}} = 1 - 2 {\cal P}  \,, \quad {\cal P}^2 = {\cal P} \,, \quad  {\cal P} \equiv  
{\cal P}({\hat x}_\mu, \theta_\alpha) \in {\mathcal A}_\theta({\mathbb R}^{1+1 \, | 2})  \otimes Mat(2) \,.
\ee
Its equation of motion is then,
\be
\frac{1}{2} (D + {\bar D}) \lbrack {\cal P} \,, (D -{\bar D}) {\cal P} \rbrack 
= \lbrack {\cal P} \,, {\bar D} D {\cal P} \rbrack = 0 \,,
\ee
and the associated conserved currents are given via the spinorial superfield
\be
{\cal J}_\alpha =  \lbrack {\cal P} \,, (D_\alpha - {\bar D}_\alpha) {\cal P} \rbrack \,.
\ee
It is instructive to present the Noether currents precisely. These are obtained through 
the $(\gamma^\mu)_{\alpha \beta} \theta_\beta$ component $j_\mu$
of ${\cal J}_\alpha$. The remaining components of ${\cal J}_\alpha$ in the Grassmann expansion, do not imply any further
conservation laws in general. Expanding ${\cal P}$ in powers of $\theta$ we have
\be
{\cal P} = P + i \theta_1 \psi_2 - i \theta_2 \psi_1 + i \theta_1 \theta_2 F \,,
\label{eq:susyproexp}
\ee
with ${\cal P}^2 = {\cal P}$ implying $P^2 = P$, $P \psi_\alpha P = 0$ and $F= i \lbrack \psi_1 \,, \psi_2 \rbrack$. 
Using (\ref{eq:susyproexp}), we find
\be
j_\mu = \lbrack P \,, \partial_\mu P \rbrack + i {\bar \psi} \gamma_\mu \psi \,.
\ee
We recognize the bosonic part as the Noether currents of the NC ${\mathbb C}P^1$ model, and the fermionic part 
as those of the NC Gross-Neveu model.

\subsection{A SUSY ${\mathbb C}P^1$ Submodel:}

We now consider a SUSY ${\mathbb C}P^1$ submodel in ${\mathcal A}_\theta({\mathbb R}^{2+1 \, | 2})$.
Extending the discussion of section 3 by including the supersymmetry, we consider the condition
\be
\lbrack {\cal P} \otimes {\cal P} \,, {\bar D} D {\cal P} \otimes {\cal P} \rbrack = 0 \,,
\label{eq:SUSYsub1}
\ee
as the defining relation for the SUSY ${\mathbb C}P^1$ submodel.

On $k$-fold tensor products $D$ is given by
\be
D = \sum_i^{k-1} \underbrace{ 1 \otimes 1 \otimes \cdots \otimes }_i
D \otimes \underbrace{ 1 \otimes 1 \cdots \otimes 1}_{k-1-j} \,.
\ee
and likewise for ${\bar D}$. We further have that
\beqa
&&({\bar D} \otimes 1 + 1 \times {\bar D} ) (D \otimes 1 + 1 \otimes D) \nn \\
&& \quad \quad = {\bar D} D \otimes  1 + {\bar D} \otimes D - D \otimes {\bar D} + 1 \otimes {\bar D} D \,,
\eeqa
the minus sign in the third term is due to the odd gradings of $D$ and ${\bar D}$.

A short calculation shows that (\ref{eq:SUSYsub1}) is equivalent to the two equations
\be
\lbrack {\cal P} \,, {\bar D} D {\cal P} \rbrack = 0 \,, \quad
{\bar D} {\cal P} \otimes \lbrack P \,, D {\cal P} \rbrack +
\lbrack {\cal P} \,, {\bar D} {\cal P} \rbrack \otimes D {\cal P} = 0 \,.
\label{eq:SUSYsub2}
\ee
Following the steps of section 3, we define
\be
{\cal J}^k_\alpha = \sum_{i=0}^{k-1} \underbrace{{\cal P} \otimes {\cal P} \cdots \otimes {\cal P}}_i 
\otimes \lbrack {\cal P} \,, (D_\alpha - {\bar D}_\alpha) {\cal P} \rbrack \otimes \underbrace{{\cal P} \otimes {\cal P} 
\cdots \otimes {\cal P}}_{k-1-i} \,.
\ee
Due to (\ref{eq:SUSYsub2}), ${\cal J}^k_\alpha$ are conserved:
\be
(D + {\bar D}) {\cal J}^k = 0 \,,
\label{eq:SUSYcons}
\ee
as can be checked explicitly for any given $k$. In components, the conserved currents are given by
\be
j^k_\mu = \sum_{i=0}^{k-1} \underbrace{P \otimes P \cdots \otimes P}_i 
\otimes \left ( \lbrack P \,, \partial_\mu P \rbrack + i {\bar \psi} \gamma_\mu \psi \right )
\otimes \underbrace{P\otimes P \cdots \otimes P}_{k-1-i} \,.
\ee
Conservation of $j^k_\mu$ is implied by the $\theta_1 \theta_2$ component of (\ref{eq:SUSYcons}).
The matrix components of $j^k_\mu$ may also be obtained using the simple procedure outlined in 
(\ref{eq:basis}), (\ref{eq:tracetensor}).

The remaining components of ${\cal J}^k_\alpha$ do not in general imply any new conservation laws.

\subsection{Solutions to the Submodel:}

The static solitonic solutions of the SUSY ${\mathbb C} P^1$ model are well known \cite{Witten-Di-Vecchia}. We can obtain 
their noncommutative versions in a straightforward manner. They are given by the BPS configurations fulfilling
\be
{\cal P} D_- {\cal P} = 0 \,, \quad \mbox{(self-dual)} \,, \quad {\cal P} D_+ {\cal P} = 0 \,, \quad \mbox{(anti-self-dual)} \,,
\label{eq:SUSYselfdual}
\ee
where the supersymmetric covariant derivatives $D_\pm = \frac{(D_1 \pm i D_2)}{\sqrt{2}}$ are given as
\footnote{In this subsection, we are using the Euclidean gamma matrices
\be
\gamma^1 = \left (
\begin{array}{cc}
1 & 0 \\
0 & -1
\end{array}
\right ) \,, \quad
\gamma^2 = \left (
\begin{array}{cc}
0 & 1 \\
1 & 0
\end{array}
\right ) \,, \quad
\gamma^5 = \gamma^1 \gamma^2 = \left (
\begin{array}{cc}
0 & 1 \\
-1 & 0
\end{array}
\right ) \,. \nn
\ee}
\be
D_+ = \partial_{\theta_-} + i \sqrt{2} \theta_- \partial_{\bar z} \,,
\quad D_- = \partial_{\theta_+} + i \sqrt{2} \theta_+ \partial_z \,,
\ee
with $\theta_\pm = \frac{\theta_1 \pm i \theta_2}{\sqrt{2}}$. They fulfil
\be
D_+^2 = i \sqrt{2} \partial_{\bar z} \,, \quad D_-^2 = i \sqrt{2} \partial_z \,, \quad \lbrace D_+ \,, D_- \rbrace = 0.
\ee

In powers of the Grassmann variables, ${\cal P}$ expands to
\be
{\cal P} = P - \theta_+ \psi _- + \theta_- \psi_+ - \theta_+ \theta_- F \,,
\ee
and ${\cal P}^2 = {\cal P}$ implies:
\be
P^2 = P \,, \quad P \psi_\pm P = 0 \,, \quad F = - \lbrack \psi_+ \,, \psi_- \rbrack \,.
\label{eq:constraints}
\ee
After using the constraints (\ref{eq:constraints}), the component form of the self-dual equation in 
(\ref{eq:SUSYselfdual}) can be cast into the equations:
\be
P \partial_z P = 0 \,, \quad P \psi_- = 0 \,, \quad P F \psi_- = 0 \,, \quad P \partial_z \psi_+ - \psi_+ \partial_z P = 0 \,.
\label{eq:componentselfdual}
\ee
From (\ref{eq:componentselfdual}) it is readily observed that the bosonic part of the solution is the BPS
solution of the NC ${\mathbb C}P^1$ model (\ref{eq:projectorlocal}). It is then easy to see that the self-dual 
solutions are given by
\be
{\cal P} = \chi \chi^\dagger \,, \quad
\chi = \left (
\begin{array}{c}
1 \\ u(z) - \theta_+ \varphi(z)
\end{array}
\right ) \frac{1}{\sqrt{u^\dagger u - \theta_+ u^\dagger \varphi - i \theta_- u \varphi^\dagger
+ i \theta_+ \theta_- \varphi^\dagger \varphi + 1}} \,, \quad \chi^\dagger \chi = 1 \,.
\label{eq:SUSYselfdualprojector}
\ee
The remaining component matrices $\psi_\pm$ and $F$ can be read off by differentiating ${\cal P}$ with respect to
$\theta_\pm$.

We can see that these configurations solve our submodel. Clearly, first of the equations in
(\ref{eq:SUSYsub2}) is automatically satisfied by the BPS equations (\ref{eq:SUSYselfdual}). As for the second
equation in (\ref{eq:SUSYsub2}), picking the self-dual configuration we have
\be
- D_+ {\cal P} \otimes D_- {\cal P} {\cal P} - D_-{\cal P} \otimes {\cal P} D_+ {\cal P}
+ {\cal P} D_+ {\cal P} \otimes D_- {\cal P} + D_-{\cal P}  {\cal P} \otimes D_+ {\cal P} \,,
\ee
which vanishes identically, after using $D_\pm  {\cal P} = D_\pm  {\cal P}  {\cal P} +  {\cal P} D_\pm {\cal P}$ together
with the self-duality equation. A similar calculation holds for the anti-self-dual case. Thus (\ref{eq:SUSYselfdualprojector})
constitute a set of solutions for the submodel under investigation.

\section{Conclusions and Outlook}

In this paper, classical integrability properties of nonlinear field theories on the Groenewold-Moyal type 
noncommutative spaces have been studied. We have obtained the infinite tower of conserved currents in the 
noncommutative principal chiral model and ${\mathbb C}P^1$ model and their supersymmetric extensions by employing 
an inductive procedure, which is well known in the corresponding commutative theories. In particular, the explicit 
expressions for the Noether currents of the noncommutative ${\mathbb C}P^1$ model, which differ from those of the 
commutative model, have been presented. We have also constructed noncommutative extensions of a ${\mathbb C}P^1$ submodel [on ${\mathcal A}_\theta({\mathbb R}^{2+1})$], as well as its SUSY extension [on ${\mathcal A}_\theta({\mathbb R}^{2+1 \, | 2})$], and proved their classical integrability by systematically obtaining their infinitely many conserved currents. In the ${\mathbb C}P^1$ submodel, a simple method to work out the explicit forms of the higher degree currents is given and it is applied on a few
examples to reveal their structure. The solitonic solutions of the submodels are also studied, and they are shown to be the same as the BPS configurations of their parent models. We think that it may be worthwhile to explore the possible connections of the ${\mathbb C}P^1$ submodel to the $U(2)$ Ward model \cite{Lechtenfeld-Popov-Chu} and their SUSY extensions. It is also interesting 
to note that there is yet another integrable ${\mathbb C}P^1$ submodel, which is defined through a weaker integrability condition
\cite{Adam1}. (Similar results in the context of the ${\mathbb C}P^N$ model in four dimensions are also known \cite{Adam2}.) It would be desirable to study its noncommutative extension as well. Progress on these topics will help us to further enhance our understanding of integrability in ${\mathcal A}_\theta({\mathbb R}^{2+1})$ and ${\mathcal A}_\theta({\mathbb R}^{2+1 \, | 2})$. We hope to report on the developments on these and related topics in the near future. 

\vskip 1em

\noindent{\bf Acknowledgements}\\[0.2em]
I thank O. Lechtenfeld for a careful reading of the manuscript and critical comments and suggestions.
This work is supported by the Deutsche Forschungsgemeinschaft (DFG) under Grant No. LE 838/9.

\end{document}